\documentstyle{article}

\def\ltap{\raisebox{-.4ex}{\rlap{$\sim$}} \raisebox{.4ex}{$<$}}
\def\gtap{\raisebox{-.4ex}{\rlap{$\sim$}} \raisebox{.4ex}{$>$}}
\newcommand{\Rsl}{{\not  \!{R}}}

\begin{document}
\vspace*{-1in}
\renewcommand{\thefootnote}{\fnsymbol{footnote}}
\begin{flushright}
SINP/TNP/99-3\\
\texttt{hep-ph/9903490} 
\end{flushright}
\vskip 5pt
\begin{center}
{\Large{\bf Effects of $R$-parity violation on direct $CP$ violation \\
in $B$ decays and extraction of $\gamma$}}
\vskip 25pt
{\sf Gautam Bhattacharyya $^{1,\!\!}$
\footnote{Electronic address: gb@tnp.saha.ernet.in}}
and 
{\sf Amitava Datta $^{2,\!\!}$
\footnote{Electronic address: adatta@juphys.ernet.in}}  
\vskip 10pt
$^1${\it Saha Institute of Nuclear Physics, 1/AF Bidhan Nagar, Calcutta 
700064, India} \\
$^2${\it Department of Physics, Jadavpur University, 
Calcutta 700032, India}
\vskip 20pt

{\bf Abstract}
\end{center}

\begin{quotation}
{\small In the standard model, direct $CP$-violating asymmetries for
$B^\pm \rightarrow \pi^\pm K$ are $\sim 2\%$ based on perturbative
calculation. Rescattering effects might enhance it to at most $\sim
(20-25)\%$. We show that lepton-number-violating couplings in
supersymmetric models without $R$-parity are capable of inducing as
large as ${\cal{O}}(100\%)$ $CP$ asymmetry in this channel. Such
effects drastically modify the allowed range of the CKM parameter
$\gamma$ arising from the combinations of the observed charged and
neutral $B$ decays in the $\pi K$ modes. With a multichannel analysis
in $B$ decays, one can either discover this exciting new physics, or
significantly improve the existing constraints on it. \\
PACS number(s): 11.30.Er, 13.25.Hw, 12.60.Jv, 11.30.Fs }
\end{quotation}

\vskip 20pt  

\setcounter{footnote}{0}
\renewcommand{\thefootnote}{\arabic{footnote}}

It is well-known that $CP$-violating $B$ decays might constitute an
important hunting ground for new physics.  This is particularly so
since many $CP$-violating asymmetries related to $B$ decays, which are
predicted to be very small in the SM \cite{sm}, are likely to be
measured with unprecedented precision in the upcoming $B$
factories. Measurements larger than the SM predictions would
definitely signal the presence of new physics. Our primary concern in
this paper is how to identify and extract such information. To
illustrate this point, let us consider direct $CP$ violation in
charged $B$ decays. The decay amplitude for $B^+\rightarrow f$ can be
written as $A(B^+\rightarrow f) =
\sum_i|A_i|e^{i\phi_i^{W}}e^{i\phi_i^{S}}$. Here $\phi_i^{W}$ and
$\phi_i^{S}$ are the weak and strong phases, respectively, for the
$i$th term. $|A_i|$ depend crucially on nonperturbative strong
interaction dynamics and have not as yet been reliably computed. One
usually measures direct $CP$ violating rate asymmetry in $B^\pm$
decays through $a_{CP} \equiv [{\cal{B}}(B^+\rightarrow f) -
{\cal{B}}(B^-\rightarrow \bar{f})]/[{\cal{B}}(B^+\rightarrow f) +
{\cal{B}}(B^-\rightarrow \bar{f})]$. Requiring $CPT$ invariance and
assuming that only two terms dominate in a given decay amplitude, the
above asymmetry can be expressed as
\begin{equation}
a_{CP} =
\frac{2|A_1||A_2|\;\sin(\phi_1^W-\phi_2^W)\;\sin(\phi_1^S-\phi_2^S)}
{|A_1|^2 + |A_2|^2 +
2|A_1||A_2|\;\cos(\phi_1^W-\phi_2^W)\;\cos(\phi_1^S-\phi_2^S)}.
\end{equation}
For $a_{CP}$ to be numerically significant the following conditions
need to be satisfied: $(i)\;|A_1|\approx |A_2|$, $(ii)\;
\sin(\phi_1^W-\phi_2^W) \approx 1$ and $(iii)\;
\sin(\phi_1^S-\phi_2^S) \approx 1$.

In the SM, the $B$ decay amplitude in a given channel receives
multiple contributions from the so-called `tree' and `penguin'
diagrams. In many cases, however, all but one of the interfering
amplitudes are highly suppressed yielding almost unobservable
$a_{CP}$. Conversely, observation of large $CP$ asymmetries in these
channels would indicate presence of amplitudes of comparable
magnitudes arising from some new physics. Consider, as an example, the
decay $B^+ \rightarrow \pi^+ K^0$. The corresponding quark level
process is $\bar{b} \rightarrow \bar{s} d\bar{d}$. In the SM, this
receives contributions only from colour-suppressed penguin
operators. Using the unitarity of the Cabibbo-Kobayashi-Maskawa (CKM)
matrix, the decay amplitude could be expressed as (following the
notations of \cite{bfm,fleischer1,fleischer2})
\begin{equation}
\label{asm}
A^{\rm SM}(B^+ \rightarrow \pi^+ K^0) =
-A\lambda^2(1-\lambda^2/2)\left[1+\rho e^{i\theta} e^{i\gamma}\right]
|P_{tc}| e^{i\delta_{tc}},
\end{equation}
where $\lambda = 0.22$ is the Wolfenstein parameter; $A \equiv
|V_{cb}|/\lambda^2 = 0.81 \pm 0.06$; $\gamma \equiv - {\rm Arg}
(V^*_{ub}V_{ud}/V^*_{cb}V_{cd})$ is the CKM weak phase; $\theta$ and
$\delta_{tc}$ are $CP$-conserving strong phases; $P_{tc} \equiv P_t^S
- P_c^S + P_t^W - P_c^W$ (the difference between top- and charm-driven
strong and electroweak penguins); and, finally, $\rho$ depends on the
dynamics of the up- and charm-penguins. For calculating $P_{tc}$ we
employ the factorization technique, which has been suggested to be
quite reliable in a recent analysis \cite{lu}. One can
express $|P_{tc}| \approx G_F f(\bar{C}_i) {\cal{F}}/\sqrt{2}$, where
${\cal{F}}=(m^2_{B_d}-m^2_{\pi})f_K F_{B\pi}$, with $F_{B\pi} = 0.3$,
and $f(\bar{C}_i)$ is an analytic function of the Wilson
coefficients\footnote{See, {\em e.g.}, Eq.~(36) of Fleischer and
Mannel \cite{flma} for the explicit dependence on the Wilson
coefficients.}. The $\bar{C}_i$'s refer to the renormalization scale
$\mu = m_b$ and denote the next-to-leading order (NLO)
scheme-independent Wilson coefficients (see the formalism developed in
Refs.~\cite{flma,buras,dh}). Assuming an average four-momentum ($k$)
of the virtual gluons and photons appearing in the penguins, we
obtain, to a good approximation, $f(\bar{C}_i) \approx 0.09$. If one
admits $0.25~\ltap~k^2/m_b^2~\ltap~0.5$, the NLO estimate of
${\cal{B}} (B^\pm \rightarrow \pi^\pm K) \equiv 0.5[{\cal{B}} (B^+
\rightarrow \pi^+ K^0) + {\cal{B}} (B^- \rightarrow \pi^- \bar{K^0})]$
varies in the range $(1.0-1.8)\times 10^{-5}$ for $\rho = 0$
\cite{flma}. This will be relevant when we subsequently confront the
experimental branching ratio with the theoretical prediction.

Following Eq.~(\ref{asm}), the $CP$ asymmetry in the $B^+\rightarrow
\pi^+ K^0$ channel is given by (neglecting tiny phase space effects)
\begin{equation} 
a_{CP}^{\rm SM} = -2\rho \sin\theta
\sin\gamma/(1+\rho^2+2\rho\cos\theta \cos\gamma).
\end{equation}
In the perturbative limit, $\rho = {\cal{O}}(\lambda^2 R_b) \sim
1.7\%$, where $R_b = |V_{ub}|/\lambda|V_{cb}| = 0.36 \pm
0.08$. However, it would be quite misleading to interpret a
measurement of $a_{CP}$ larger than a few $\%$ as a signal of new
physics. Rescattering effects
\cite{rescatter,bfm,fleischer1,fleischer2}, such as, $B^+ \rightarrow
\pi^0 K^+ \rightarrow \pi^+ K^0$, {\em i.e.}, long-distance
contributions to the up- and charm-driven penguins, could enhance
$\rho$ to as large as ${\cal{O}}(10\%)$, based on an
order-of-magnitude estimate using Regge phenomenology. On account of
such final state interactions, $a_{CP}$ could jack up to
${\cal{O}}(20\%)$, which is non-negligible. It has recently been shown
that new physics enhanced colour dipole coupling and destructive
interference could push $a_{CP}$ to as large as $\sim 40\%$
\cite{hou}. What if a much larger $a_{CP}$ is observed?

In the minimal supersymmetric standard model, there are additional
contributions to the $B^\pm \rightarrow \pi^\pm K$ penguins, and
$a_{CP}$ could go up to $\sim 30\%$ \cite{barbieri}. But switching on
$R$-parity-violating ($\Rsl$) interactions \cite{rpar,review} triggers
topologically different new diagrams and their interference with the
SM penguins could generate large $a_{CP}$. Its quantitative estimate,
as far as practicable, is the thrust of the present paper. The key
point is that the lepton-number-violating interactions of the type
$\lambda'_{ijk} L_i Q_j D^c_k$, which is a part of the $\Rsl$
superpotential, could contribute to non-leptonic $B$ decays at the
tree level. For some of these channels the leading SM contributions
arise from penguins. In view of the current upper bounds on the
relevant $\lambda'$ couplings and sparticle masses \cite{review}, the
possibility that the $\Rsl$ tree contributions are of the same order
of magnitudes as the SM penguins is very much open. Moreover, $\Rsl$
interactions are potentially important sources of new weak phases. To
clarify this point, we first stress that the $\lambda'_{ijk}$ are in
general complex. Even if a given $\lambda'$ is predicted to be real in
some specific model, the phase rotations of the left- and right-handed
quark fields required to keep the mass terms real and reduce the CKM
matrix to its standard form will automatically introduce a new weak
phase in this term, barring accidental cancellations. Indeed, this
phase can be absorbed by redefining the slepton or the sneutrino
field. But the $\Rsl$ contribution to a $B$ decay amplitude depends on
the product of the type $\lambda'_{ij3}\lambda'_{ilm}$. This product
cannot be rendered real by transforming the single slepton or
sneutrino field corresponding to the index $i$. This point has not
been emphasized in the literature.

How $R$-parity violation induces $CP$ violation in neutral $B$ decays
have been analysed before \cite{kaplan,guetta,jang}. In this paper, we
examine the $\Rsl$ effects on direct $CP$ violation in $B^\pm
\rightarrow \pi^\pm K$ decays. We show that these effects could cast a
much larger numerical impact than the non-perturbative dynamics ({\em
e.g.}, the rescattering effects) in the SM. In the process we have
derived new upper bounds on the $\lambda'_{i13}\lambda'_{i12}$
combinations from the existing data. We also study how the same $\Rsl$
couplings contaminate the extraction of the CKM phase $\gamma$ from $B
\rightarrow \pi K$ decays. These issues have not been addressed in the
past.

How to compute the $\Rsl$ contributions to the decay $B^+ \rightarrow
\pi^+ K^0$?  One can generate a colour non-suppressed tree level
amplitude for this process using the simultaneous presence of the
$\lambda'_{i13}$ and $\lambda'_{i12}$ couplings. Following the
standard practice, we shall assume that only one such pair (for a
given $i$) of $\Rsl$ couplings is numerically significant at a
time. This constitutes a sneutrino ($\tilde{\nu}_i$) mediated
decay. Using a simple Fierz transformation, one can rearrange this
$(S-P)(S+P)$ interaction in a $(V-A)(V+A)$ form. The amplitude of this
new contribution turns out to be\footnote{In Ref.~\cite{huitu}, a
correction factor ($f_{\rm QCD} \simeq
(\alpha_S(m_b)/\alpha_S(\tilde{m}))^{24/23} \simeq 2$ for a sneutrino
mass ($\tilde{m}$) of 100 GeV) to such $\Rsl$ interactions has been
computed. We do not indulge ourselves with this nitty-gritty for our
order of magnitude estimate.}
\begin{equation}
\label{arsl}
A^{\Rsl}(B^+ \rightarrow \pi^+ K^0) = (|\lambda'_{i13}\lambda'^*_{i12}|/8
\tilde{m}^2){\cal{F}} e^{i\gamma_{\Rsl}} \equiv -|\Lambda_{\Rsl}|
e^{i\gamma_{\Rsl}},
\end{equation}
where $\gamma_{\Rsl}$ is the weak phase associated with the product of
$\lambda'$s. The total amplitude (sum of expressions in
Eqs.~(\ref{asm}) and (\ref{arsl})) can be parametrized as
\begin{equation} 
A(B^+ \rightarrow \pi^+ K^0) = -A\lambda^2(1-\lambda^2/2) |P_{tc}|
e^{i\delta_{tc}} \left(1+\rho e^{i\gamma} e^{i\theta} + \rho_{\Rsl}
e^{i\gamma_{\Rsl}} e^{i\theta_{\Rsl}}\right),
\end{equation} 
where 
\begin{equation}
\rho_{\Rsl} \equiv |\Lambda_{\Rsl}|/A\lambda^2(1-\lambda^2/2)
|P_{tc}|.
\end{equation}
Note that $\rho_{\Rsl}$ is free from uncertainties due to factorization.
A straightforward computation for the net $CP$ asymmetry yields
\begin{equation}
\label{acp}
a_{CP} = -\frac{2\rho \sin\theta \sin\gamma + 2\rho_{\Rsl}
\sin\theta_{\Rsl} \sin\gamma_{\Rsl} + 2 \rho \rho_{\Rsl} \sin(\theta -
\theta_{\Rsl}) \sin(\gamma - \gamma_{\Rsl})}{1 + \rho^2 +
\rho_{\Rsl}^2 + 2\rho \cos\theta \cos\gamma + 2\rho_{\Rsl}
\cos\theta_{\Rsl} \cos\gamma_{\Rsl} + 2 \rho \rho_{\Rsl} \cos(\theta -
\theta_{\Rsl}) \cos(\gamma - \gamma_{\Rsl})}. 
\end{equation}
It is clear that $a_{CP}$ is numerically insignificant if both $\rho$
and $\rho_{\Rsl}$ are vanishingly small. As an illustrative example to
demonstrate that $R$-parity violation alone has the potential to
generate a $CP$ asymmetry much larger than the most optimistic
expectation in the SM, we set $\rho = 0$ in Eq.~(\ref{acp}). A
non-zero $\rho (\sim 0.1)$ could dilute the effect at most by $\sim
20\%$.

At this point, an estimate of how large $\rho_{\Rsl}$ could be is in
order. We choose $\tilde{m} =$ 100 GeV throughout our analysis.
Employing the current upper limits on $\lambda'_{i13}\lambda'_{i12}$
\cite{review}, we obtain, for $i =$ 1, 2, and 3, $\rho_{\Rsl}~\ltap$
0.17, 3.45, and 4.13 respectively\footnote{These upper limits
correspond to the products of the individual upper limits obtained
from different physical processes assuming a common superpartner
scalar mass of 100 GeV \cite{review}.}. The upshot is that it is
possible to arrange $\rho_{\Rsl} = 1$ (for $i =$ 2, 3), which implies
that a 100\% $CP$ asymmetry is no longer unattainable, once we set
$\gamma_{\Rsl} = \theta_{\Rsl} = \pi/2$. Such a drastic hike of $CP$
asymmetry is somewhat unique and we emphasize that it is hard to find
such large effects in other places \cite{chemtob}.  Notice that a
minimum $\rho_{\Rsl}$ is required to generate a given $a_{CP}$. This
is given by (for $\rho = 0$)
\begin{equation}
\label{rhomin}
\rho_{\Rsl} ~\gtap ~(1-\sqrt{1-a_{CP}^2})/|a_{CP}|.
\end{equation}
Eq.~(\ref{rhomin}) has been obtained by minimizing $\rho_{\Rsl}$ with
respect to $\gamma_{\Rsl}$ and $\theta_{\Rsl}$ for a given
$a_{CP}$. Numerically,
\begin{equation}
\label{rhonum}
\rho_{\Rsl}~\gtap~ 1.0~(1.0), 0.50~(0.8), 0.33~(0.6), 0.21~(0.4),
0.10~(0.2);
\end{equation}
where the numbers within brackets refer to the corresponding
$a_{CP}$'s.

At the same time we must ensure that $\rho_{\Rsl}$ does not become too
large to overshoot the experimental constraint on ${\cal{B}} (B^\pm
\rightarrow \pi^\pm K)$. The latest CLEO measurement reads
${\cal{B}}^{\rm exp}(B^\pm \rightarrow \pi^\pm K) = (1.4
\pm 0.5 \pm 0.2)\times 10^{-5}$, which means ${\cal{B}}^{\rm exp} \leq
1.9\times 10^{-5}$ (1$\sigma$) \cite{cleo}. Recall that
${\cal{B}}^{\rm SM} \sim (1.0 - 1.8)\times 10^{-5}$ in view of the
present uncertainty of the SM (for $\rho = 0$; varying $\rho$ within
$0-0.1$ cannot change ${\cal{B}}^{\rm SM}$ significantly). Therefore
the tolerance to accommodate a multiplicative new physics effect could
at most be a factor of 1.9 (at 1$\sigma$). It is easy to extract from
the denominator of Eq.~(\ref{acp}) that pure $\Rsl$ effects modify the
SM prediction of the branching ratio by a multiplicative factor $(1 +
\rho^2_{\Rsl} + 2\rho_{\Rsl} \cos\theta_{\Rsl}
\cos\gamma_{\Rsl})$. The maximum allowed value of $\rho_{\Rsl}$ is
then obtained by setting one of the two angles appearing in this
factor to zero and the other to $\pi$ ({\em i.e.}, arranging maximum
possible destructive interference in the branching ratio). This leads
to the conservative upper bound
\begin{equation} 
\label{rhoup} 
\rho_{\Rsl} ~\ltap ~2.4;~~{\rm implying}~~|\lambda'_{i13}\lambda'^*_{i12}|
~\ltap ~5.7\times 10^{-3}~~(1\sigma).
\end{equation}
Note that, in spite of the large uncertainties as discussed above, for
$i = 2, 3$, these limits are already stronger than the existing ones
(see discussion just above Eq.~(\ref{rhomin})). Moreover, the
existing bounds from semi-leptonic processes necessarily depend on the
exchanged squark masses and have been derived by assuming a common
mass of 100 GeV for them. With present data from Tevatron on the
squarks and gluino searches, this seems to be a too optimistic
assumption. On the other hand, our bounds from hadronic $B$ decays are
based on a more realistic assumption that the virtual sneutrinos have
a common mass of 100 GeV. We also note that the choice of phases that
leads to Eq.~(\ref{rhoup}) kills the $\Rsl$ contribution to the $CP$
asymmetry. If, on the other hand, we are interested in finding the
upper limit on $\rho_{\Rsl}$, with $a_{CP}$ maximized with respect to
$\gamma_{\Rsl}$ and $\theta_{\Rsl}$, we must set each of the angles to
$\pi/2$: the interference term vanishes, and we obtain a stronger
limit $\rho_{\Rsl} ~\ltap ~1.0$. This, in conjunction with
Eqs.~(\ref{rhonum}) and (\ref{rhoup}), defines a range of the $\Rsl$
couplings to be probed in the upcoming $B$ factories.

The $\Rsl$ couplings responsible for such large $CP$ asymmetries might
leave important side-effects that could surface out while extracting
the angles ($\alpha$, $\beta$ and $\gamma$) of the unitarity triangle.
In the SM, the three angles measured independently should sum up to
$\pi$.  In the presence of new physics, if one attempts to
determine those angles assuming the validity of the SM alone, one may
not obtain their true values. The SM amplitude of $B_d \rightarrow
\Psi K_s$, regarded as the gold-plated channel for extracting $\beta$,
is dominated by a sum of operators all multiplying the same CKM
combination $V^*_{cs} V_{cb}$. It so happens that the present bounds
on the corresponding $\Rsl$ couplings, {\em viz.}  $\lambda'_{i23}$
and $\lambda'_{i22}$, are too strong to contaminate the extraction of
$\beta$ in a numerically significant manner ($A^{\Rsl} (B_d
\rightarrow \Psi K_s)~\ltap~ 0.02 A^{\rm SM} (B_d \rightarrow \Psi
K_s)$) \cite{kaplan,guetta}. On the other hand, extraction of $\alpha$
{\em via} $B_d \rightarrow \pi^+\pi^-$ may be contaminated in a
significant way, once $\lambda'_{i13}\lambda'_{i11}$ combinations are
switched on. However, in the scenario that we are considering ({\em
i.e.}, where only $\lambda'_{i13}\lambda'_{i12}$ combinations are
non-zero), extractions of both $\alpha$ and $\beta$ remain
unaffected. Having thus measured $\alpha$ and $\beta$, $\gamma$ can be
determined indirectly using the relation $\gamma = \pi - \alpha -
\beta$.

Let us now consider a direct measurement of $\gamma$, as suggested in
Ref.~\cite{flma}, where one uses the observable $ R \equiv
[{\cal{B}}(B_d \rightarrow \pi^-K^+) + {\cal{B}}(\bar{B}_d \rightarrow
\pi^+K^-)]/ [{\cal{B}}(B^+\rightarrow \pi^+ K^0) +
{\cal{B}}(B^-\rightarrow \pi^- \bar{K^0})].  $ The present
experimental range is $R = 1.0 \pm 0.46$ \cite{cleo}. Neglecting
rescattering effects, a measurement of $R$ could be used to bound
$\gamma$ within the framework of the SM as $\sin^2\gamma ~\ltap ~R$
\cite{flma}. Within errors, it may still be possible that $R$ settles
to a value significantly smaller than unity, disfavouring values of
$\gamma$ around $90^\circ$. This will certainly be in conflict if, for
example, $\gamma \approx 90^\circ$ is preferred by indirect
determination.

Now considering the same $\Rsl$ scenario expressed through
Eq.~(\ref{arsl}), the SM bound is modified to ($\rho = 0$)
\begin{equation}
\label{gbound3}
\sin^2\gamma ~\ltap~ R~(1 + \rho^2_{\Rsl} + 2\rho_{\Rsl}
\cos\theta_{\Rsl} \cos\gamma_{\Rsl}).
\end{equation}
Thus the bound on $\gamma$ either gets relaxed or further constrained
depending on the magnitude of $\rho_\Rsl$ and the signs of
$\gamma_{\Rsl}$ and $\theta_{\Rsl}$. For $\rho_\Rsl \approx 1$ and
$\gamma_{\Rsl} = \theta_{\Rsl} \approx \pi/2$, $\sin^2\gamma ~\ltap~
2R$, and hence there is no constraint on $\gamma$, if $R$ turns out to
be $\gtap$ 0.5. Notice that with these choices of $\rho_\Rsl$,
$\gamma_{\Rsl}$ and $\theta_{\Rsl}$, one expects to observe large $CP$
asymmetry as well. Therefore the lesson is that if a large $a_{CP}$ is
observed, more care is necessary in extracting $\gamma$. But the
latter in isolation cannot provide a comprehensive signal of new physics.

The existing bound on $\gamma$ in the SM ($41^\circ ~\ltap~ \gamma~
\ltap~ 134^\circ$) has been obtained from a global fit of the
unitarity triangle using data on $|V_{cb}|$, $|V_{ub}|/|V_{cb}|$,
$B_d$--$\bar{B}_d$ mixing and $CP$ violation in the $K$ system
\cite{bound}. The allowed zone is almost symmetric around $\gamma =
90^\circ$. Interestingly, a measurement of $R < 1$, which is still
possible within errors, excludes a region symmetric with respect to
$\gamma = 90^\circ$. Eq.~(\ref{gbound3}) implies that these two
contrasting features can be reconciled by $R$-parity violation.

It would be worth performing a multichannel analysis combining all
kinds of $B \rightarrow \pi \pi, \pi K $ \cite{agashe} and $B
\rightarrow D K$ modes \cite{dk} (recommended for measuring $\gamma$)
to enhance the significance of any nonzero asymmetry and to identify
the sources of new physics in a more unambiguous way. In this context,
$R$-parity violation has a distinctive feature that it can enhance
both $a_{CP}$ and the branching ratios simultaneously. With our choice
of $\Rsl$ couplings, only $B^\pm \rightarrow \pi^\pm K$ modes are
affected.  Note that if we turn on the $\lambda'_{i23}$ and
$\lambda'_{i11}$ couplings instead of those we have considered here,
we obtain the same diagrams except that the new amplitude picks up a
colour-suppression factor. On the other hand, it is possible to probe
$\Rsl$ effects on other channels, {\em e.g.}, $B_s\rightarrow \phi K$
($b \rightarrow \bar{s}ss$), by turning on other $\Rsl$
couplings. Even if no enhanced $a_{CP}$ is observed, or no disparity
between ${\cal{B}}^{\rm SM}$ and ${\cal{B}}^{\rm exp}$ is established,
new constraints on $\Rsl$ couplings could be obtained, as has already
been hinted by Eq.~(\ref{rhoup}). All these signify that a systematic
study of many observables together with different sets of $\Rsl$
couplings could constitute an exciting program in view of the upcoming
$B$ factories.

The work of AD has been supported by the DST, India (Project No.
SP/S2/k01/97) and the BRNS, India (Project No. 37/4/97 - R \& D
II/474). Both authors thank A. Raychaudhuri for a careful reading of
the manuscript, and M. Artuso of the CLEO Collaboration for
communication regarding new data. 


\end{document}